\documentclass[twocolumn,showpacs,preprintnumbers,amsmath,amssymb]{revtex4}

\makeatletter

\usepackage{graphicx}
\usepackage{dcolumn}
\usepackage{bm}

\newcommand{\La}{\Lambda}

\newcommand{\Ne}{{N_\mathrm{e}}}
\newcommand{\Nc}{{N_\mathrm{c}}}

\newcommand{\calC}{\mathcal{C}}
\newcommand{\calB}{\mathcal{B}}

\newcommand{\sigmaG}{\sigma_\mathrm{G}}
\newcommand{\SG}{\mathcal{S}_\mathrm{G}}
\newcommand{\cs}[1]{c_{#1,\sigma}}
\newcommand{\csd}[1]{c_{#1,\sigma}^\dagger}

\newcommand{\as}[1]{\alpha_{#1,\sigma}}
\newcommand{\asd}[1]{\alpha_{#1,\sigma}^\dagger}

\newcommand{\aps}[1]{\alpha_{#1,\sigma}^\prime}
\newcommand{\apsd}[1]{\alpha_{#1,\sigma}^{\prime\dagger}}

\newcommand{\bs}[1]{\beta_{#1,\sigma}}
\newcommand{\bsd}[1]{\beta_{#1,\sigma}^\dagger}

\newcommand{\ns}[1]{n_{#1,\sigma}}
\newcommand{\nt}[1]{n_{#1,\tau}}

\begin{document}

\title{Physical realization of the four color problem 
in quantum systems}

\author{Masanori Yamanaka$^1$ and Akinori Tanaka$^2$}

\affiliation{$^1$Department of Physics, College of Science and Technology,
Nihon University, Kanda-Surugadai 1-8-14, Chiyoda-ku,
Tokyo 101-8308, Japan}
\affiliation{$^2$Department of General Education,
Ariake Natonal College of Technology, Omuta, 
Fukuoka 812-8581, Japan
}

\date{\today}

\begin{abstract}
A multi-component electron model on a lattice is constructed
whose ground state exhibits a spontaneous ordering 
which follows the rule of map-coloring used
in the solution of the four color problem.
The number of components is determined
by the Euler characteristics of a certain surface
into which the lattice is embedded.
Combining the concept of chromatic polynomials
with the Heawood-Ringel-Youngs theorem,
we derive an index theorem relating the degeneracy 
of the ground state with a hidden topology of the lattice.
The system exhibits coloring transition 
and hidden-topological structure transition.
The coloring phase exhibits a topological order.
\end{abstract}

\pacs{71.10.Fd, 75.10.Nr, 64.60.-i, 02.30.Ik}

\maketitle

Physics and Mathematics are two disciplines in Exact Science which 
are known to nurture each other. There are famous problems 
in Mathematics which are 
physically realizable. An example is how to determine 
the configuration of 
$N$ repulsive point particles confined inside a spherical shell.
It is then natural to ask 
whether there is a possible realization of 
other famous mathematical challenges.
Among them we focus here on the celebrated four-color problem, 
formulated by F.~Guthrie in 1852~\cite{REF:4cbib}.
This problem remained unsolved for more than hundred years
until an affirmative answer was proved by
Appel and Haken~\cite{REF:fourcolour}. 
Alongside, a plethora of new mathematical concepts were introduced,
such as a computer-aided proof 
and Non-deterministic Polynomial-time (NP)-hard problems.
In the framework of chromatic problems in Mathematics,
coloring is a {\it passive} procedure to the map.
Namely, the colors are artificially assigned to the regions
by try-and-error so as to satisfy mathematical rules.
A more focused question is whether there is an {\it active} coloring 
as a phenomenon in Nature,
by which we mean a spontaneous emergence of an ordering
such as molecular crystallization, magnetization, and orbital,
whose ordering pattern follows the map-coloring rule.

In this work we give an affirmative answer to this question 
in terms of an electron model with several components (colors)
defined on an arbitrary lattice. 
The pertinent ground state exhibits spontaneous ordering 
which follows the map-coloring rule. 
(Although the Potts model is known to exhibit
a similar state~\cite{REF:Wu},
the mechanism is trivial due to the anitiferromagnetic 
exchange interaction between classical spin variables.
While ours is a quantum mechanical one.) 
Employing the concept of
chromatic polynomials and combining it
with the Heawood-Ringel-Youngs theorem
~\cite{REF:Heawood,REF:4cbib},
we derive an index theorem, relating the degeneracy 
of the electronic ground state with a 
hidden topology of the random lattice.
Conversely, for a given number of components,
we propose a constructive method for building a lattice
on which the exact ground state can be realized.
To this end we modify 
the Haj\'os construction~\cite{REF:Hajos,REF:4cbib}
of a random graph.
The ground state exhibits coloring transition
where the coloring phase is characterized by a topological order.
This electronic model, hence, provides a meeting point 
for condensed matter physics, manifold embedding 
and topological graph theory. 
It is also another route to a realization 
of a topological order
~\cite{REF:to1,REF:to11,REF:to12,REF:to2,REF:to3,REF:to4,REF:to5,REF:to6} 
in quantum system.

The model Hamiltonian (see below)
is composed of local non-negative operators.
The procedure of constructing its ground state
is very similar to the one employed in other models 
~\cite{REF:mg,REF:AKLT}.
Electronic realization of such Hamiltonian is presented in
Refs.~\cite{REF:Brandt,REF:Tasaki,REF:Tasaki2}.
Here we consider a multi-component model.
The number of components, {\it e.g.} spin or orbital projections,
is denoted by $N_c$.

\noindent
\begin{figure}[b]
\includegraphics[width=80mm]{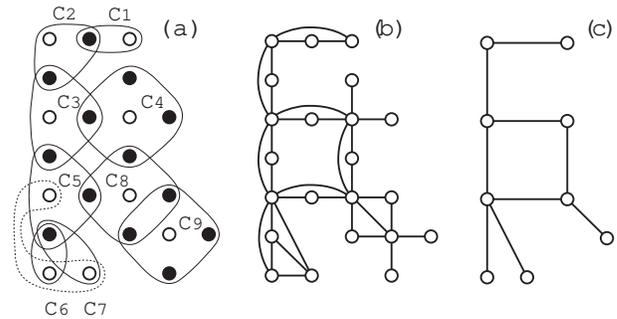}
\caption{(a) Example of sites and cells. 
The cells, $C_1$, $C_2$, $C_3$, and $C_4$
are composed of two, three, four, and five sites respectively. 
The open and bold circles denote the capital $i\in\calC$
and border $u\in\calB$, respectively.
A border can belong to several different cells.
(See $C_5$, $C_6$, and $C_7$.)
Two cells can share several different borders.
(See $C_8$ and $C_9$.)
$D_u$ for the site shared by $C_5$, $C_6$, and $C_7$
is shown by broken line.
(b) The lattice associated with (a). 
The lines denote possible hopping of electron.
(c) The dual graph associated with (a).
}
\label{FIG:celldual}
\end{figure}

To expose the complete Hamiltonian,
the ``cell construction'' technique~\cite{REF:Tasaki,REF:Tasaki2} 
is employed.
Consider the usual finite random lattice sites
$\Lambda$, which can be written as $\La=\calC\cup\calB$
where $\calC\cap\calB=\emptyset$.
(The usual regular and periodic lattices are allowed.)
For each site $i\in\calC$, we define a \textit{cell} 
$C_i=\{i\}\cup B_i$ where $B_i$ is a subset of $\calB$ 
with $|B_i|\ge1$~\cite{REF:commentX}.
(See Fig.~\ref{FIG:celldual}(a).)
The cells $C_i$ and $C_j$ are said to be directly connected 
if they share at least one site
(i.e., $C_i\cap C_j\ne\emptyset$).   
We assume that every site in $\calB$ is contained in at least one
cell and, for any pair of cells $C_i$ and $C_j$, 
there is a pass of directly connected cells {from} $C_i$ to $C_j$.
A given site $u\in\calB$ may be shared by several cells
and the number of cells to which the site $u$
belongs is the coordination number of cell, $N_u$.
We assume that $max(N_u) \le N_c$.
As usual, we denote by $\cs{x}(\csd{x})$  
the fermion annihilation (creation) operator
at the site $x\in\La$ with the component $\sigma$, and introduce
fermion operators,
\begin{eqnarray}
 \as{i}&=&\lambda_{i}^{(i)}\cs{i}
 +\sum_{u\in B_i}\lambda_{i}^{(u)}\cs{u} \ \ 
 {\rm for} \ \ i\in\calC 
 \label{eq:alpha}
 \\
 \bs{u}&=&\mu_u^{(u)}\cs{u}
 +\sum_{i\in D_u} \mu_{u}^{(i)}\cs{i} \ \ 
 {\rm for} \ \ u\in\calB,
\end{eqnarray}
where $D_u$ is the set of sites $i$ in $\calC$ such that $u\in C_i$.
(See Fig.~\ref{FIG:celldual}(a).)
The $\as{i}^{\dagger}$ and $\bs{u}^{\dagger}$ 
create the localized single-electron states with the component $\sigma$
in the cell $C_i$ and $\{u\}\cup D_u$, respectively.
Here, we assume that the real parameters $\lambda_{i}^{(x)}$ and
$\mu_{u}^{(x)}$ are nonvanishing and they satisfy 
$\lambda_{i}^{(i)}\mu_{u}^{(i)}+\lambda_{i}^{(u)}\mu_{u}^{(u)}=0$
for any pair of $i(\in\calC)$ and $u(\in\calB)$ contained 
in the same cell.
We consider the Hamiltonian
\begin{eqnarray}
H&=&\sum_{\sigma=1}^{N_c}\sum_{u\in\calB}\bsd{u}\bs{u}
+U\sum_{i\in\calC}\sum_{\sigma,\tau=1;\sigma>\tau}^{\Nc}\ns{i}\nt{i}
\nonumber\\
&& +V\sum_{i\in\calC}\sum_{u\in B_i}\sum_{\sigma=1}^{\Nc}\ns{i}\ns{u}, 
\label{eq:hamiltonian}
\end{eqnarray}
where $n_{i,\sigma}=c_{i,\sigma}^{\dagger}c_{i,\sigma}$ 
is the number operator, and $U>0$ and $V>0$.
The first term is the hopping term (see Fig.~\ref{FIG:celldual}(b)), 
the second one is the onsite Coulomb repulsion,
and the last one is the density correlation.
For $\Nc=2$ with $\sigma=(\uparrow, \downarrow)$
the system describes fermions with spin 1/2.
To construct a model with $f$ orbitals one set $N_c=2f$.
With the further assumption such that
$\lambda_{i}^{(u)}$ are positive for any $i\in\calC$ and $u\in B_i$,
we can also consider another Hamiltonian 
\begin{eqnarray}
H&=&\sum_{\sigma=1}^{N_c}\sum_{u\in\calB}\bsd{u}\bs{u}
  +U\sum_{i\in\calC}\sum_{\sigma,\tau=1;\sigma>\tau}^{\Nc}
                \ns{i}\nt{i}
\nonumber\\
&&+V\sum_{i\in\calC}\sum_{\sigma=1}^{\Nc}\ns{i}\apsd{i}\aps{i}
\label{eq:hamiltonian2}
\end{eqnarray}
where $\aps{i} = \sum_{u\in B_i}\cs{i}$.

In the following we fix the total electron number
$\Ne=\sum_{i, \sigma}n_{i, \sigma}$ to $|\calC|$
and find the exact ground states of $H$.
Consider the states of the form
\begin{equation}
 \Phi(\{\sigma\})
 =\prod_{i\in\calC}\alpha_{i,\sigma_i}^\dagger \vert 0 \rangle,
\end{equation}
where $\{\sigma\}$ represents 
a configuration of components $(\sigma_i)_{i\in\calC}$ 
of the $\alpha$-operators (\ref{eq:alpha}).
Let $\{\sigmaG\}$ be a configuration 
satisfying that $\sigma_i\ne\sigma_j$ 
if the cells $C_i$ and $C_j$ are directly connected,
and denote by $\SG$ the collection of these configurations.
Note that a site in $\calB$ may be occupied by several 
(not more than $N_c$) electrons.
Using the anticommutation relation $\{\asd{i},\bs{u}\}=0$
for $i\in\calC$ and $u\in\calB$,
we find that the state $\Phi(\{\sigmaG\})$ is a zero mode 
of the Hamiltonians (\ref{eq:hamiltonian}) and (\ref{eq:hamiltonian2}) 
and satisfies the relation $H \Phi(\{\sigmaG\})=0$.
It is evident that the Hamiltonians are 
non-negative (positive semidefinite) and hence, 
$\Phi(\{\sigmaG\})$ is an exact ground state.
It is possible to prove that any ground state can be
represented as
\begin{equation}
 \sum_{\{\sigmaG\}\in\SG}\phi(\{\sigmaG\})\Phi(\{\sigmaG\})
\end{equation}
with coefficients $\phi(\{\sigmaG\})$.

\begin{figure}[b]
\includegraphics[width=80mm]{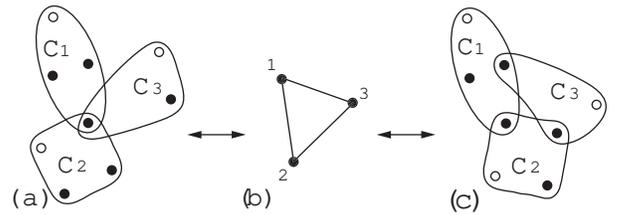}
\caption{Two types of a configuration of cells (a) and (c),
whose dual graph is the same (b).
Then, for a given dual graph (b), we have two types of
local configuration of the cells.
The coordination number of the cell, $N_u$, of the site
which is shared by $C_1$, $C_2$, and $C_3$ in (a) is 3.
The number of the site 
which is shared by $C_1$, $C_2$ in (c) is 2.
}
\label{FIG:redund}
\end{figure}

In order to elaborate upon the ground state 
and relation to the coloring problem, let us fix the terminology.
A {\it dual graph} $G(\Lambda)$ is associated 
with the lattice $\Lambda$ in such a way that
a {\it vertex} of  $G(\Lambda)$ is 
defined by a cell $i$ of $\Lambda$.
When the cells $C_i$ and $C_j$ are directly connected,
an {\it edge} in $G(\La)$ is defined 
between the vertices $i$ and $j$. 
(See Figs.~\ref{FIG:celldual}(a) and (c).)
Note carefully that 
the {\it dual graph} so defined
is different {from} the usual {\it dual lattice}
associated with the original lattice $\Lambda$.
In standard Graph Theory, 
the lattice and its dual lattice
have a one-to-one correspondence.
In the present case,
the dual graph is defined uniquely for a given lattice, 
but the inverse is not necessarily true.
Due to this property, we have two kinds of redundancy
as the lattice is constructed {from} a given dual graph.
(See Fig.~\ref{FIG:redund}.)
(i) The number of sites composing a cell is arbitrary.
(ii) The coordination number of cell, $N_u$, is arbitrary even
when the coordination number of vertices is fixed.
This plays a crucial role in the modification 
of the Haj\'os construction. As we argue below, 
the existence problem of the ground state 
is now transformed into that
of vertex coloring of the dual graph.

\noindent

For the case where the dual graph is planar,
we can regard the cell $C_i$ as the country
and the component $\sigma_i$ as its color.
(In this case, the site $i$ is the capital and the sites 
in $B_i$ shared by several cells are the border lines.) 
The rule for generating configurations $\{\sigmaG\}$ 
is equivalent to the map-coloring rule.
Therefore we call $\{\sigmaG\}$ as {\it coloring}.  
The answer to the question of whether a 
coloring $\{\sigmaG \}$ does exist depends on $N_c$ 
and the structure of the lattice.
Finding coloring for a given lattice 
is recognized as one of the NP-hard problems.
Fortunately, in the present coloring problem,
there are a couple of routes to approach it. The first one 
tells us
how to look for $N_c$ for a given lattice. Here
the four color and Heawood-Ringel-Youngs theorems 
are applicable respectively for the two-dimensional planar 
graph and for higher-dimensional random graphs.
The second route guides us
how to construct the lattice for a given $N_c$.
Here we need to modify the Haj\'os 
construction~\cite{REF:Hajos,REF:4cbib}
of random lattice.

As a first step, consider a lattice $\La$ whose dual graph 
is planar.
Employing the four color theorem~\cite{REF:fourcolour},
any random planar graph is at least four colorable,
and there exists at least one ground state
of the model with $N_c \ge 4$. 
For the model with $N_c < 4$,
existence of a ground state depends on the structure of the lattice.
For example, the ground state on a lattice
whose dual graph is bipartite
can be constructed for $N_c \ge 2$
because the bipartite graph is two-colorable.
When a dual graph is $N_c$-colorable but not 
$(N_c-1)$-colorable, we refer to $N_c$
as the {\it chromatic number for constructing the ground state}
and denote it by $N_c^*$.
(Note that if a graph is $N_c$-colorable, 
it is also colorable 
for the color $>$ $N_c$~\cite{REF:4cbib}.)

Consider now a general random lattice.
The problem of finding $\SG$ is then equivalent 
to coloring the random hyper-polyhedra.
The dual graph $G$ is in general not planar 
and the four color theorem does not apply.
Arbitrary dual graphs can be embedded 
into a higher surface with a proper genus $g$~\cite{REF:4cbib}.
They are classified into 
the torus with the Euler characteristic $\chi=2-2g$,
the projective plane with $\chi=1-2g$, 
and the Klein bottle with $\chi=-2g$~\cite{REF:twofold}. 
We respectively denote them by $T^g$, $P^g$, and $K^g$.
Now recall the
Heawood-Ringel-Youngs theorem
~\cite{REF:Heawood,REF:4cbib}.
Let $G$ be a dual graph on a higher surface $S$ 
of Euler characteristic $\chi$, and let each cell 
of the lattice have at most $m$ disjoint connected parts.
Except for $(m, \chi)=(1, 2)$,
an upper bound on $N_c^*$ is given by 
\begin{eqnarray}
N_c=\Bigg[ \frac{6m+1+\sqrt{(6m+1)^2-24\chi}}{2} \Bigg],
\label{eq:heawood} 
\end{eqnarray}
where $[ \ \ ]$ is the Gaussian symbol.

Conversely, for a given $N_c$, 
the lattice with $N_c^*=N_c$ is obtained 
by proper recursive use of the following procedures (i), (ii) and (iii): 
(i) Prepare the complete $N_c$-graph, consisting of $N_c$
vertices and edges connecting any two of vertices.
Replace each vertex by an arbitrary cell
and identify their sites in such a way 
that the lattice so constructed has the complete $N_c$-graph
as its dual graph.
(ii) In a dual graph of an already obtained lattice,
identify two vertices not jointed by an edge
and form a new dual graph.
(ii-a) Replace two cells associated with the two vertices identified above
by a proper (arbitrary) cell, and identify sites 
so that the resulting lattice maintains the connectivity
of the new dual graph.
(ii-b) If the coordination number of cell, $N_u$, of the sites 
exceeds $N_c$,
reduce it by the relation shown 
in Fig.\ref{FIG:redund}(a)$\to$(b)$\to$(c).
If any such reduction attempt fails, increase the number 
of sites in the replaced cell,
or go back to (ii) and choose another pair of vertices 
for identification. 
(iii) (modification of the Haj\'os construction): 
For disjoint dual graphs $G_1$ and $G_2$
of lattices which are already obtained,
remove an edge between the vertices $x_1$ and $y_1$ in $G_1$
(which means the separation of two cells $C_{x_1}$ and $C_{y_1}$),
and remove an edge between $x_2$ and $y_2$ in $G_2$.
Identify $x_1$ and $x_2$, and add the edge between $y_1$ and $y_2$,
(which means an identification of site in the cells $C_{y_1}$ 
and $C_{y_2}$).
Create a lattice by following the same ways as (ii-a) and (ii-b).
  
Generically, the ground state is degenerate.
The degree of degeneracy, $\Gamma$, is equivalent 
to that of the different coloring possibilities 
of the dual graph $G(\Lambda)$.
In the coloring problem, 
the number of the different coloring possibilities of a graph $G$
is described by the chromatic polynomial $P(G,x)$
where $x$ is the number of colors~\cite{REF:4cbib}. 
Therefore, we obtain the number of zero modes, 
\begin{eqnarray} 
\Gamma = P\big(G(\Lambda), N_c\big).
\label{eq:chropoly}
\end{eqnarray} 
For some categories of graphs, the chromatic polynomials are known.
For a general graph, however, 
derivation of its chromatic polynomial is an NP-hard problem.
Our analysis then shows that the degree of degeneracy is determined 
through (\ref{eq:heawood}) and (\ref{eq:chropoly}).
This index theorem is quite intriguing,
because the algebraic and topological structures of the higher surface
determine the properties of the interacting electron model.

\begin{figure}[t]
\includegraphics[width=80mm]{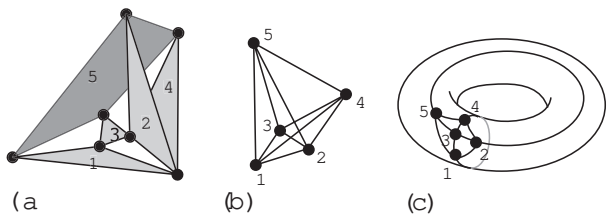}
\caption{
(a) Lattice composed of five cells.
Only the borders which are shared by more than two cells are shown.
(b) Dual graph associated with the lattice.
(c) Irreducible embedding of the dual graph into 
the two-dimensional manifold. 
}
\label{FIG:complete5}
\end{figure}

As an example, we demonstrate the above procedures
within a simple lattice. (See Fig.~\ref{FIG:complete5}(a).)
The dual graph is shown in Fig.~\ref{FIG:complete5}(b)
which is termed as $K_5$, {\it the complete graph of degree five}.
$K_5$ is irreducibly embedded into $T^2$ 
whose Euler characteristics is $\chi=2-2\times 1=0$.
{From} eq.~(\ref{eq:heawood}), we obtain $N_c=7$,
which is an upper bound of $N_c^*$.
The chromatic polynomial for $K_5$ is known~\cite{REF:4cbib} as 
\begin{eqnarray} 
P(K_5, x) = \prod_{p=1}^{5}(x-p+1).
\end{eqnarray} 
The $N_c^*$ is given 
by the minimal integer
which satisfies $P(K_5, x) > 0$, that is, $x=5$. 
Hence, for a given $N_c \ge 5$ 
the number of zero modes is given by
$\Gamma= \prod_{p=1}^{5}(N_c-p+1)$.
An extension of this procedure to the other lattices 
is straightforward.

\noindent

We now discuss two types of exotic phase transitions
which this system exhibits. Consider an arbitrary lattice
with $N_c^*$. 
Denote by ${\cal M}$ the manifold into which the dual graph 
is embedded irreducibly. 

The first type is a 
{\it coloring transition}~\cite{REF:colortrans}
(For a context of the Potts model,
see~\cite{REF:colortrans21,REF:colortrans22,REF:colortrans23,
REF:colortrans24,REF:colortrans25}.)
Assume that the number of components is controllable
by injection and absorption.
For example, when the components are related to electron spin,
one can control the number of components
by applying a magnetic field. 
And when the components are related to an orbital degree of freedom,
it is controlled by a lattice distorsion.
For $N_c > N_c^*$,
the ground state is given by $\Phi(\{\sigmaG\})$
and the system is in a coloring phase.
If we reduce the number of components 
and obtain a system with $(N_c-1)$ components
on the same lattice, 
the ground state is also described 
by $\Phi(\{\sigmaG\})$ with $(N_c-1)$ components.
Within this operation,
the degeneracy of the ground 
state undergoes a {\it macroscopic} change.
The corresponding change of the system entropy
is accompanied
by an emission or absorption of latent heat.
When the number of components is reduced further,
the coloring transition occurs 
at $N_c=N_c^*$
because 
for $N_c < N_c^*$ 
the ground state is not described by $\Phi(\{\sigmaG\})$.
As the system goes through $N_c^*$ the pattern 
of orbital ordering should undergo a dramatic change.
(Mathematically, we define the transition point 
by the smallest zero of the chromatic polynomial 
associated with the dual graph.)
The coloring phase, $N_c > N_c^*$,
is characterized by a global consistency of coloring.
In this sense, the ground state has a topological order.

The second type is a
{\it hidden-topological structure transition}.
This is  a novel structural transition realized 
for a fixed number of components
and lattice sites $\La$.
In this transition there is a variation 
of the manifold ${\cal M}$ 
into another manifold ${\cal M}'$
which undergoes a modification of the hopping network.
The transition is characterized 
by the set of manifolds $({\cal M}, {\cal M}')$
and, as before, is accompanied by a 
transfer of latent heat.
For a given random lattice, 
the ground state properties are essentially determined 
by the topological structure of the higher surface
into which the dual graph is embedded irreducibly, and
not by the original lattice structure in real space
(hence the adjective ``hidden'').
The ground state is then classified 
by the hidden topology.

Future research directions should include
detailed studies pertaining to numerous
concrete lattices and their transitions.
The case of infinite lattices is of special interest 
in view of the thermodynamic limit 
and its relation to Ramsey theories.

Acknowledgments: We would like to thank 
Yshai Avishai and Akihiro Tanaka 
for critical reading of the manuscript.


\begin{thebibliography}{99}
\bibitem{REF:commentX}
We use the symbol $|X|$ to denote the number of elements 
in a set $X$.
\end{thebibliography}
\end{document}